\documentstyle[psfig,times]{mn}

\def\etal{et~al.}
\def\spose#1{\hbox to 0pt{#1\hss}}
\def\lta{\mathrel{\spose{\lower 3pt\hbox{$\mathchar"218$}}
     \raise 2.0pt\hbox{$\mathchar"13C$}}}
\def\gta{\mathrel{\spose{\lower 3pt\hbox{$\mathchar"218$}}
     \raise 2.0pt\hbox{$\mathchar"13E$}}}

\title[The environments of radio--loud AGN in the 2dFGRS]{The environmental
dependence of radio--loud AGN activity and star formation in the 2dFGRS}

\author[P.~N.~Best]{P.~N.~Best,$^1$\thanks{Email:
pnb@roe.ac.uk}\\ 
$^1$ Institute for Astronomy, Royal Observatory Edinburgh, Blackford Hill,
Edinburgh EH9 3HJ, UK\\
}

\pagerange{\pageref{firstpage}--\pageref{lastpage}}
\pubyear{1999}
\begin{document}
\label{firstpage}

\maketitle

\begin{abstract}
\noindent By combining the 2--degree Field Galaxy Redshift Survey with the
NRAO VLA Sky Survey at 1.4\,GHz, the environments of radio loud AGN in the
nearby Universe are investigated using both local projected galaxy
densities and a friends--of--friends group finding algorithm. Radio--loud
AGN are preferentially located in galaxy groups and poor--to--moderate
richness galaxy clusters. The AGN fraction appears to depend more strongly
on the large--scale environment (group, cluster, etc) in which a galaxy is
located than on its more local environment, except at the lowest galaxy
surface densities where practically no radio--loud AGN are found. The ratio
of absorption--line to emission--line AGN changes dramatically with
environment, with essentially all radio--loud AGN in rich environments
showing no emission lines. This result could be connected with the lack of
cool gas in cluster galaxies, and may have important consequences for
analyses of optically--selected AGN, which are invariably selected on
emission line properties. The local galaxy surface density of the
absorption--line AGN is strongly correlated with radio luminosity, implying
that the radio luminosities may be significantly boosted in dense
environments due to confinement by the hot intracluster gas.

The environments of a radio--selected sample of star forming galaxies are
also investigated to provide an independent test of optical studies. In
line with those studies, the fraction of star forming galaxies is found to
decrease strongly with increasing local galaxy surface density; this
correlation extends across the whole range of galaxy surface densities,
with no evidence for the density threshold found in some optical studies.
\end{abstract}


\section{Introduction}

The advent of large galaxy redshift surveys, especially the 2-degree Field
Galaxy Redshift Survey (2dFGRS; Colless \etal\ 2001)\nocite{col01} and the
Sloan Digital Sky Survey (SDSS; York \etal\ 2000; Stoughton \etal\
2002)\nocite{yor00,sto02} have revolutionised our understanding of the
effect of local environment upon the evolution of galaxies. Understanding
how galaxy properties, such as luminosities, morphologies, star formation
rates and nuclear activity, depend upon the environment that a galaxy
inhabits can place important constraints on models of galaxy formation and
evolution, and allow the intrinsic properties of the galaxies to be
separated from those that have been externally induced (`nature {\it vs}
nurture').

It has been known for many years that star formation rates are strongly
suppressed in the central regions of galaxy clusters (e.g. Dressler et~al
1985),\nocite{dre85} even if account is taken of the different distribution
of morphological types in cluster environments as compared with the
field. The large redshift surveys have shown that this suppression of the
star formation rate is not only restricted to the extreme cluster
environments, but begins at much lower environmental densities.  Hashimoto
\etal\ \shortcite{has98} showed that the mean star formation rate shows a
continuous correlation with local galaxy density, both inside and outside
of clusters, using the Las Campanas Redshift Survey. Lewis et~al (2002;
hereafter Lew02)\nocite{Lew02} studied the fields around 17 clusters within
the 2dFGRS, and found that the mean star formation rate of galaxies is
relatively constant for projected galaxy surface densities below 1 galaxy
per square Mpc, but that at higher surface densities star formation is
increasingly suppressed, down to essentially zero at 50 galaxies per square
Mpc. A similar study using SDSS data broadly supports these conclusions
although with a weaker break (G{\'omez \etal\ 2003; hereafter
Gom03)\nocite{Gom03}, and a study of SDSS field galaxies (Mateus \&
S{\'o}dre 2003; hereafter MS03)\nocite{mat03} is also in agreement at high
surface densities, but suggests that the star formation rate remains
sensitive to the local galaxy density even in more rarefied environments.

Regardless of the precise dependence at lower surface densities,
environmental effects clearly function down to at least 1 galaxy per square
Mpc, which is below both the mean galaxy surface density at the virial
radius of relatively rich clusters and that of galaxy groups. Therefore,
the physical processes that lead to this quenching of star formation are
not intrinsic to cluster environments (e.g. ram--pressure stripping of the
interstellar medium by the hot cluster gas; Gunn \& Gott
1972),\nocite{gun72} but also occur in smaller structures
(cf. Mart{\'\i}nez \etal\ 2002, who find that star formation is diminished
even within galaxy groups of mass $M \sim 10^{13}$ in the
2dFGRS).\nocite{mar02b} The combination of this environmental dependence
and the build up of galaxies into groups and clusters with cosmic time may
be one of the drivers behind the decline in the mean cosmic star formation
rate since redshifts $z \sim 1$ \cite{mad98}.

An alternative handle on galaxy activity and environmental influence comes
through studies of active galactic nuclei (AGN). It is now apparent that
essentially all massive galaxies in the nearby Universe host a supermassive
black hole at their centres, whose mass is roughly proportional to the
spheroidal mass of the galaxy (e.g. see review by Kormendy \& Gebhardt
2001)\nocite{kor01}. This suggests that the build-up of the central black
hole and that of its host galaxy are fundamentally linked; the similarity
of the cosmic evolution of the mean global star formation rate to that of
the rate at which gas is accreted onto black holes (as estimated from the
radio luminosity function; Dunlop 1998),\nocite{dun98} at least out to
redshifts $z \sim 1$-2 where both are well-determined, provides further
evidence for this. Investigating how the incidence of AGN activity depends
upon local environment can provide valuable insight into the origin of this
connection, and also into the physical processes that trigger AGN activity.

To produce a powerful AGN the necessary ingredients are a massive black
hole, a supply of fuel, and a transport mechanism to connect the two. If
AGN activity is driven predominantly by the availability of the cold gas
(the same factor that drives star formation activity) then AGN activity,
like star formation, should be greatly suppressed in cluster
environments. If instead the probability of AGN activity is solely a
function of the central black hole, independent of the local environment of
the galaxy, the AGN fraction would simply trace the distribution of galaxy
bulges. Alternatively, at least for the most powerful AGN it has frequently
been proposed that galaxy interactions or mergers may trigger the AGN
activity (e.g. Hutchings \& Campbell 1983, Heckman \etal\ 1984,
1986),\nocite{hut83,hec84,hec86} providing both a supply of gas and a
mechanism for moving it to the central regions of the galaxy. In this case
the dependence of AGN activity on environment on group-- or cluster--scales
would be less clear--cut, with AGN favouring those environments in which
conditions are optimal for galaxy interactions and mergers.

Studies of the environmental dependence of AGN activity have a long and
chequered history. Dressler \etal\ \shortcite{dre85} argued that AGN
activity was suppressed in clusters, finding that only 1\% of cluster
galaxies showed AGN activity compared to 5\% of field galaxies in their
sample.\footnote{Note that the fraction of galaxies which host AGN is very
dependent upon both the depth of the observations and the details of the
spectral definition of an AGN: Ho \etal\ \shortcite{ho97} classify as many
as 43\% of the galaxies in their survey of nearby bright ($B_T < 12.5$)
galaxies as active.} However, the redshift range of their cluster and field
samples were very different, and the field sample contained many higher
redshift AGN which were only bright enough to make it into their sample
because of the magnitude boosting effect of the AGN itself. Biviano \etal\
\shortcite{biv97} find that if a correction is applied for this magnitude
bias then the lower incidence of AGN activity in clusters is consistent
with being due solely to the difference in the morphological mix between
cluster and field. Carter \etal\ \shortcite{car01} similarly find little
evidence for an environmental dependence of AGN fraction.

Using the 2dFGRS and the SDSS it is possible to go beyond a binary
comparison of cluster against field and investigate the environmental
dependence of AGN activity in much greater detail, as has been done for
star formation activity. Kauffmann et~al \shortcite{kau03} robustly
selected a sample of AGN from the SDSS survey, by modelling the underlying
stellar continuum to obtain accurate measurements of the Balmer emission
lines, and then using emission line ratio diagnostics to separate the AGN
from star forming galaxies. Kauffmann et al (in prep; private
communication) find that these AGN, especially those with strong emission
lines, are less common in regions of high galaxy surface density. However,
Miller et~al \shortcite{mil03a}, also using the SDSS, found that the
fraction of galaxies hosting AGN remains roughly constant across all
environments from cluster cores to the rarefied field.

The contradiction between these two sets of results indicates that
selection of active galaxies by optical emission lines is neither
straightforward nor uncontroversial. In addition, recent Chandra X-ray
observations of galaxy clusters have identified a population of X-ray
active cluster galaxies which would not be selected as active galaxies on
the basis of either their optical or emission line properties (e.g. Martini
\etal\ 2002)\nocite{mar02a}, and radio studies \cite{mil02a} have similarly
found a population of dust--obscured active and star--forming galaxies
towards the central regions of clusters. Therefore, selecting active
galaxies by means other than their emission line properties may provide
more robust results, and will certainly provide a valuable test of the
results currently being derived from the SDSS. An efficient way to do this
is to use radio-loud AGN: these are straightforward to locate and study,
and large--area deep radio surveys are already available. It must be borne
in mind, however, that only 10-20\% of AGN are radio loud (e.g. Hooper
\etal\ 1995)\nocite{hoo95}, and these may represent a biased subset of the
AGN population as a whole.

The environments of powerful radio sources have been widely studied at low
redshifts (e.g. Prestage \& Peacock 1988; Hill \& Lilly 1991; Miller \etal\
2002)\nocite{pre88,hil91,mil02b} and appear to favour galaxy groups and
weak clusters; they tend to avoid the densest environments except in the
few cases where they are in the special location of being hosted by the
central dominant galaxy of a cluster. To date, though, these studies have
been both limited to the most powerful radio sources and, generally, based
upon galaxy number count statistics or cross--correlation analyses, with
little or no spectroscopic redshift information (the Miller \etal\ study is
the exception to this, providing redshifts for a handful of galaxies in the
$\sim 20$ arcmin radius region around each of 25 low redshift radio
sources). By comparing the 2dFGRS with a deep large--area radio survey such
as the NRAO VLA Sky Survey (NVSS; Condon \etal\ 1998)\nocite{con98} it is
possible to overcome both of these deficiencies, studying the
spectroscopically--determined large--scale environments of `typical'
radio--loud AGN within the 2dFGRS, and comparing these to those of the
galaxy population in general. This is the goal of the current paper.

In Section~\ref{samples}, the galaxy and radio source samples used in the
analysis are defined. The methods used to derive the properties of these
galaxies are described in Section~\ref{methods}. The results of the
environmental analysis are described in Section~\ref{results}, and these
are discussed in Section~\ref{discuss}. Section~\ref{concs} summarises the
results. Throughout the paper, the values adopted for the cosmological
parameters are $\Omega_m = 0.3$, $\Omega_{\Lambda} = 0.7$, and $H_0 =
65$\,km\,s$^{-1}$Mpc$^{-1}$.

\section{Galaxy samples and properties}
\label{samples}

\subsection{2dFGRS galaxy sample}
\label{2dfsamp}

The primary galaxy sample for this study is drawn from the 2dF Galaxy
Redshift Survey, described by Colless \etal\ \shortcite{col01}.  The 2dFGRS
obtained spectra through $\sim$2.1 arcsec diameter fibres for nearly a
quarter of a million galaxies brighter than a nominal extinction-corrected
magnitude limit of $b_J=19.45$, in two declination strips (one equatorial
and the other close to the south galactic pole; these are hereafter
referred to as the NGP and SGP strips respectively) and a number of random
2--degree diameter fields. For this paper, analysis was restricted to those
galaxies with reliable redshifts (quality $\ge 3$; cf. Colless \etal\ 2001)
which lie within the NGP or SGP strip.

A cut in redshift was made to select those galaxies which lay in the
redshift range $0.02 < z < 0.1$. At redshifts below $z \sim 0.02$ the
physical size of the 2dFGRS survey region is too small to investigate the
large--scale galaxy environments, whilst the upper redshift cut was set by
the depth of the 2dFGRS survey, as discussed below. The absolute B--band
magnitudes of these galaxies were calculated, using the average
K--correction for 2dFGRS galaxies derived by Madgwick \etal\
\shortcite{mad02}; $K_B(z) \approx 1.9z + 2.7z^2$. Only those galaxies with
absolute magnitudes $M_{B} < -19$ (that is, $L_{B} \gta 0.25 L_{B}^*$; cf
Norberg et~al 2002)\nocite{nor02} were retained for subsequent analysis;
this absolute magnitude limit corresponds roughly to the apparent magnitude
limit of the survey at a redshift $z \sim 0.1$. Thus, the combination of
absolute magnitude and redshift cuts largely removes any redshift biases
from the sample. 

This produced a sample of 56143 galaxies. However, not all of these
galaxies are appropriate for environmental analysis, for example because
they lie close to the boundaries of the survey, or in regions of low
spectroscopic completeness. This was assessed for each galaxy individually
during the process of local environment estimation through the 10th nearest
neighbour method, as discussed in Section~\ref{10near}: about 10\% of the
galaxies were rejected from the sample during this analysis, due to their
locations. Note that removal of these galaxies will not bias any of the
results of the paper, firstly because these are largely a random subset of
galaxies which happen to lie near the boundaries of the survey, and
secondly because in any case these same cuts are applied to all of the
samples under study.

By this process, a basis catalogue of 50684 galaxies with redshifts $0.02
\le z \le 0.10$ and absolute magnitudes $M_B < -19$ was derived, for each
of which good estimates of the local galaxy density can be made.

\subsection{Radio source sample}
\label{radsamp}

From this basis catalogue, subsamples of radio galaxies are constructed
using the radio source sample of Sadler \etal\ (2002; hereafter
Sad02)\nocite{sad02}. These authors cross-correlated an earlier version of
the 2dFGRS, which contained almost 30\% of the final 2dFGRS catalogue, with
the radio sources in the NRAO VLA Sky Survey \cite{con98}, which is a
1.4\,GHz survey to a limiting flux density of about 2.5\,mJy covering the
entirety of the sky north of $-$40 degrees declination. Sad02 identified
all of the single component radio sources which lay within 15 arcseconds of
a galaxy with a redshift in the 2dFGRS catalogue, and also added to this
sample a small number of double or multiple component radio sources for
which an examination of a radio--optical overlay indicated a host galaxy
within the 2dFGRS catalogue. For each of the 912 candidate radio source
matches, they examined the optical emission and absorption line spectrum of
the host galaxy, to classify the source as:

\noindent (i) `star-forming' galaxies (SF)\footnote{Star forming galaxies
emit at radio wavelengths, predominantly due to the synchrotron emission of
particles accelerated in supernova shocks. The radio luminosity is roughly
correlated with the star formation rate: a 1.4\,GHz radio luminosity of
$10^{22}$\,W\,Hz$^{-1}$ corresponds to a star formation rate of order $5
M_{\odot}$\,yr$^{-1}$ (e.g. Condon 1992 and references therein; Carilli
2001)\nocite{con92,car01b}.}, where the spectrum is dominated by strong narrow
Balmer emission lines.

\noindent (ii) AGN, separated into three subclasses: absorption--line
ellipticals galaxies (Aa), absorption line galaxies with weak LINER--like
emission lines (Aae), and galaxies with spectra dominated by nebular
emission lines such as [OII] or [OIII] (Ae).

\noindent (iii) sources whose classifications are likely (SF?, Aa?, Ae?,
Aae?) or completely uncertain (???). 

\noindent These classifications were made by visual examination of the
spectra, and their reliability was checked against Principle Component
Analyses and emission line diagnostic diagrams
(Sad02)\nocite{sad02}.\footnote{Note that because the 2dFGRS spectra are
uncalibrated, it is not possible to classify the galaxies in a more robust
way, such as decomposing the stellar continuum to investigate the presence
of a young stellar population, or modelling and subtracting the stellar
continuum to prevent stellar absorption lines from affecting the emission
line properties (cf. Kauffmann \etal\ 2003).\nocite{kau03}}

For the current analysis, stricter criteria need to be applied to these
radio sources, in order to prevent false identifications from unduly
influencing the results. Firstly, following the procedure of Sad02, the
radio source sample is limited to only those radio sources with
radio--optical offsets below 10 arcseconds, or in the case of resolved or
multi--component radio sources with larger radio--optical offsets, only
where the visual examination shows the identification to be secure.  This
restricts the size of the overall sample to 757 radio sources, of which up
to 10\% may be chance coincidences (Sad02)\nocite{sad02}. The radio source
sample is further restricted to include only those galaxies in the final
2dFGRS sample constructed above, namely by restricting it to the redshift
range $0.02 \le z \le 0.10$, to galaxies with absolute magnitudes $M_B <
-19$, and to galaxies for which local environments were able to be
calculated, as discussed above. A final restriction was made to include
only radio sources with radio luminosities\footnote{A radio spectral index
of $\alpha = 0.8$ (for $S_\nu \propto \nu^{-\alpha}$) was adopted for the
K--corrections in calculating the rest--frame radio luminosities.} above
$10^{21}$W\,Hz$^{-1}$ at 1.4\,GHz rest--frame. This provided a final sample
of 272 radio sources with optical counterparts. Of these, the proportion of
chance coincidences is $\lta 5$\%. This is reduced from the 10\% in the
original sample because of the various cuts applied to the data, especially
the restriction to the most luminous optical galaxies; this is because the
probability of an optical galaxy being a powerful radio source is an
increasing function of optical luminosity (e.g. Sadler \etal\
1989)\nocite{sad89}, and so the removal of the lowest optical luminosity
galaxies removes a significant fraction of the chance alignments but very
few of the genuine sources.

It is important that the star forming and AGN radio source subsamples are
cleanly selected from these 272 radio sources. Naively assuming that all
sources classified by Sad02 as `SF?' can be adopted as star forming
galaxies, and all `Ae?', `Aa?' and `Aae?' as AGN, may lead to some
mis-classification which could strongly bias the results. The radio
luminosity was therefore used to provide a second handle on the source
classification. The majority of low luminosity radio sources are star
forming while at high luminosities AGN dominate (e.g. Sad02; Miller \& Owen
2002; Machalski \& Godlowski 2000)\nocite{sad02,mil02a,mac00}. The
transition between the two classes is gradual and the exact location at
which it occurs is still a matter of debate with Sad02 favouring $L_{\rm
1.4\,GHz} \approx 10^{23}$W\,Hz$^{-1}$ and Miller \& Owen favouring the
lower value of $10^{22.7}$W\,Hz$^{-1}$. This difference may arise because
the Miller \& Owen study concentrates upon cluster environments and, as
discussed in the introduction, star formation activity is known to be
suppressed there, decreasing the proportion of star forming galaxies.

A luminosity of $L_{\rm 1.4\,GHz} \approx 10^{22.8}$W\,Hz$^{-1}$ is here
adopted for the transition point. The subsample of star forming galaxies is
taken to include all of these galaxies securely classified by Sad02 as star
forming galaxies (`SF'), plus those galaxies classified as likely star
forming galaxies (`SF?') which also have a radio luminosity below this
transition value (that is, likely star forming galaxies based upon their
emission lines, and also more likely to be star forming based upon their
radio luminosities). 154 radio sources satisfy these criteria, with
1.4\,GHz radio luminosities ranging from $3 \times 10^{21}$ to $4 \times
10^{23}$W\,Hz$^{-1}$. The AGN subsample is taken to include those sources
securely classified as AGN by Sad02 (classes Aa, Ae, and Aae), together
with those sources classified as likely AGN (Aa?, Ae?, Aae?) with radio
luminosities above the cut-off; there are 91 sources thus classified as
AGN, with radio luminosities ranging from $10^{22}$ to $1.4 \times
10^{25}$W\,Hz$^{-1}$. This additional radio luminosity selection may not
avoid all mis-classifications, but does greatly reduce the likelihood of
them. It naturally leads to the exclusion of a small subset of perfectly
good radio sources (`SF?' sources with 1.4\,GHz radio luminosities brighter
than $10^{22.8}$\,W\,Hz$^{-1}$, and `Aa?', `Ae?' and `Aae?' sources fainter
than that luminosity) but their exclusion is unlikely to bias the results
significantly, and it is considered preferable to work with a smaller but
cleaner sample than to risk including mis-classified objects.

It is finally important to note that the classification method of Sad02
takes no account of the recent results of Kauffmann \etal\
\shortcite{kau03}, that many AGN are found to be composite systems which
also contain star formation activity. Since a fibre instrument such as the
2dF samples a large fraction of the galaxy light, the star forming
component is likely to dominate the emission line properties of such
composite objects, and they are more likely to be classified as star
forming galaxies.  There may therefore be some composite SF + AGN systems
found within the SF class: however, the relatively lower radio luminosities
of the SF galaxies (together with the substantially different environmental
properties found in this paper) suggest that such objects are relatively
rare. It is plausible that many of the 27 objects rejected as having
uncertain classifications are composite objects. 

Two further samples of radio sources were also defined: (i) Luminous AGN,
defined to be the subset of the AGN class which also had $L_{\rm 1.4\,GHz}
\ge 10^{23}$W\,Hz$^{-1}$, and (ii) Extended radio sources, defined as the
subset of AGN for which the NVSS radio emission was extended. These two
subsamples contain 40 and 25 radio sources respectively. 21 of the 25
extended radio sources also satisfy the luminous AGN definition, and so the
extended radio sources are largely, but not entirely, a subset of the
luminous AGN class.
  
\subsection{Parent galaxy sample}
\label{optsamp}

As discussed above, the Sad02 radio sample was not constructed from the
entirety of the 2dFGRS survey, but from only the first $\sim 30$\%
observed. Ideally the radio source samples defined from the Sad02 regions
should therefore be compared with the parent galaxy sample used by Sad02,
and not with the entire 2dFGRS region, since otherwise there is the
possibility of introducing biases if the initial regions studied by the
2dFGRS were not fully representative of the complete survey
region. However, the parent catalogue of Sad02 is not available.  Whether
the Sad02 parent sample is biased with respect to the final survey
catalogue can be investigated by comparing the final 2dFGRS galaxy
catalogue with the subsample of galaxies that were included in the first
catalogue release of the 2dFGRS (100,000 galaxies, of which the Sad02
sample comprised 70\%). Results of the environmental analyses discussed in
Section~\ref{methods} were compared for the final 2dFGRS galaxy sample and
the first 100,000 galaxy subsample. One notable difference was found
between the two samples: a significantly smaller percentage of the galaxies
in the 100,000 galaxy catalogue are found the very highest local densities
(richest clusters) than in the final release catalogue. Further
investigation shows that this difference arises predominantly because of a
single supercluster environment (at RA 13h, Dec -2$^{\circ}$, $z \sim
0.084$) which was very poorly covered in the original data release.

Apart from this one extreme environment, no systematic differences were
found between the two samples in either local densities or
group\,/\,cluster environments. Therefore, given that the Sad02 parent
sample comprises $\sim 70$\% of the 100,000 galaxy release, and that the
100,000 galaxy release contains no especially extreme environments, any
biases between the Sad02 parent sample and the first--release 100,000
galaxy sample are likely to be negligible. Therefore, the parent galaxy
sample adopted for comparison with the radio source samples comprises the
21085 galaxies from the 50684-galaxy basis catalogue described above, which
were present in the first 100,000 galaxy release of the 2dFGRS.

\subsection{Morphological sample}
\label{sdsssamp}

The SDSS First Data Release \cite{aba03} overlaps in part with the NGP
strip of the 2dFGRS. For those galaxies in the overlap region, their
morphological properties can be investigated using the SDSS `concentration
parameter', C. This parameter is defined as the ratio of the radius
enclosing 90\% of the galaxy light in the $r$-band to that containing 50\%
of the light (cf. Stoughton et~al. 2002)\nocite{sto02}: most early--type
galaxies have $C > 2.6$ while spirals and irregulars typically have $2.0 <
C < 2.6$ \cite{str01}.

For the 50684 galaxies in the final 2dFGRS sample, a search was made within
the SDSS catalogue for any galaxy within 5 arcsec in positional offset
($\approx 10$\,kpc at redshift 0.1), and within 340\,km\,s$^{-1}$ in
redshift (four-times the rms deviation of 85\,km\,s$^{-1}$ between SDSS and
2dFGRS measured redshifts; cf Norberg et~al 2002).\nocite{nor02} The
completeness and reliability of this cross--correlation were estimated by
varying the size of the acceptable offsets in both velocity and position,
and comparing both the total number of 2dFGRS--SDSS matches and the number
of incidences whereby two SDSS galaxies are consistent with the 2dFGRS
galaxy; it is estimated that for allowable offsets of 5 arcsec and
340\,km\,s$^{-1}$ the cross--correlation of the two surveys is about 98\%
complete and 99.9\% reliable.

For all galaxy matches, the concentration parameter C was calculated from
the SDSS $r$-band data, and the galaxy was classified as either early or
late--type. Although this was only possible for 5781 galaxies, this is
sufficient to determine the morphological mix of 2dFGRS galaxies as a
function of environment.

\section{Environmental Parameters}
\label{methods}

The local environment of each galaxy in the survey was investigated using
two different methods, which are described below.

\subsection{10th-nearest neighbour estimate}
\label{10near}

For each galaxy with redshift $0.02 \le z \le 0.10$ and absolute magnitude
$M_B < -19$, the local projected galaxy density was calculated using an
adaptation of the 10th-nearest neighbour approach which has been commonly
used for this purpose (cf. Dressler \etal\ 1980, Lew02, Gom03, Miller
\etal\ 2003).\nocite{dre80,Lew02,Gom03,mil03a} Specifically, for each
galaxy, a redshift shell of $\pm 1000$\,km\,s$^{-1}$ centred on the
redshift of that galaxy is considered, and the projected distance
($r_{10}$) to the 10th nearest neighbour [also having $M_B < -19$] within
that redshift shell in the basis catalogue is calculated. This distance is
then converted into a local projected galaxy density, $\Sigma = 10 / \pi
r_{10}^2$.

Several factors deserve deeper consideration in this determination of the
surface densities. The first is the adoption of a fixed velocity range of
1000\,km\,s$^{-1}$ in the definition of the redshift shell, to reject
foreground and background galaxies. This velocity range is well--suited to
`average' environments, but in the richest clusters the large velocity
dispersions may lead this method to underestimate the local galaxy density,
due to some companions falling outside of this range. However, this is
unavoidable without making the velocity shell so wide as to include
unacceptable numbers of foreground and background galaxies. Lew02
\nocite{Lew02} got around this problem by limiting their analysis to
cluster environments, and using a variable--width velocity shell set to be
three times the velocity dispersion of each cluster. However, this method
does not allow analysis of field environments, limiting the range of
surface densities that can be studied, and in addition it may introduce
biases at the lowest surface densities (see discussion in
Section~\ref{sfdiscuss}).  An alternative method for investigating the
environments of field galaxies is to convert redshifts into distances and
calculate the full three--dimensional local density of the galaxies
(e.g. see MS03)\nocite{mat03}; however, peculiar velocities make this
method very poor in cluster environments. Overall the adoption of a fixed
velocity range provides a method that works reliably across all
environments, with sufficient accuracy for the science goals of this paper.
The reliability of this technique was further tested by varying both the
width of the velocity shell ($\pm 500$, 1000, 1500\,km\,s$^{-1}$) and the
number of neighbours out to which the result was evaluated (5, 10, 15):
although the projected surface densities of individual galaxies changed,
the overall sample--average results were broadly unaffected.

A second effect concerns the boundaries of the 2dFGRS catalogue, and the
variations in both redshift completeness and magnitude limit of the
catalogue as a function of position. The redshift survey covers two long
strips in declination, each of which is only about 10 degrees wide, meaning
that all galaxies lie within 5 degrees of a catalogue boundary, and many
lie significantly closer. To estimate the effect that this will have on the
surface density calculation, the following procedure was followed:

\noindent (i) For each galaxy, the value of $r_{10}$ was calculated as
above. 10000 positions were then randomly chosen within a circle of radius
$r_{10}$ centred on that galaxy.

\noindent (ii) For each of these 10000 positions, the redshift completeness
of the 2dFGRS catalogue was evaluated for galaxies brighter than the
apparent magnitude corresponding to an absolute magnitude of $M_{B} = -19$
at the redshift of the galaxy under study. This calculation was carried out
using the redshift completeness, limiting magnitude, and $\mu$-masks (see
Colless \etal\ 2001 for details) provided by the 2dFGRS team.  If a given
position lay outside the 2dFGRS region of study, or within one of the
survey holes associated with bright stars, then the redshift completeness
was set to zero for that position.

\noindent (iii) An average redshift completeness was derived for these
10000 positions. In this way, for each galaxy the average redshift
completeness for galaxies brighter than $M_{B} < -19$ over the region of
sky out to radius $r_{10}$ was estimated.

\noindent (iv) Where the average redshift completeness was below 50\%, or
the value of $r_{10}$ was in excess of 5 degrees, the galaxy was removed
from further analysis; these are predominantly galaxies near the boundaries
of the survey region. Removal of these galaxies will not bias any of the
results of the paper, since these same cuts are applied to all of the
samples under study.

\noindent (v) The surface densities calculated for the remainder of the
galaxies were scaled up by the inverse of their average redshift
completeness. This correction factor provides a rough first--order
approximation to the correction and, since all applied corrections are
below a factor of 2, any inaccuracies will have a negligible effect on the
final results.

\subsection{Group or cluster membership}
\label{gpclus}

The 10th nearest neighbour analysis technique was developed at a time
before large spectroscopic catalogues were available, and foreground and
background field galaxies had to be statistically subtracted. The use of
redshift information to remove clearly unrelated galaxies greatly improves
the reliability of the analysis, and there are clearly many advantages over
simple comparisons with, for example, cluster-centric radius. However, this
technique 
still discards some of the information which is available in a
complete spectroscopic dataset, especially in galaxy groups or clusters
with over 10 members. Using the velocity information available in the
2dFGRS and SDSS catalogues, it is in principle possible to determine for
each individual galaxy its exact environment: is it an isolated galaxy, a
member of a small group, on the outskirts of a cluster, within a cluster
core, etc?

Catalogues of galaxy groups have been constructed from the 2dFGRS by both
Merch{\'a}n \& Zandivarez \shortcite{mer02} and by Norberg \etal\
\shortcite{nor03}, in both cases using an adapted version of the {\it
friends--of--friends} approach first described by Huchra \& Geller
\shortcite{huc82}. The basis of this approach is as follows. Two galaxies
with an angular separation on the sky of $\theta_{12}$, and recession
velocities of $V_1$ and $V_2$ (with a mean, $V = (V_1 + V_2)/2$) are
considered to be linked if they satisfy the following conditions:

\begin{displaymath}
2 {\rm sin}\left(\frac{\theta_{12}}{2}\right) \frac{V}{H_{0}} \le D_{\rm L},
\end{displaymath}

\begin{displaymath}
|V_1 - V_2| \le V_{\rm L}
\end{displaymath}

\noindent where $D_{\rm L}$ and $V_{\rm L}$ are the transverse and radial
linking lengths respectively. For a survey of fixed apparent magnitude
limit, in the standard Huchra \& Geller formalism the linking lengths
$D_{\rm L}$ and $V_{\rm L}$ vary with recession velocity in order to
compensate the change in the sampling of the galaxy luminosity function
with distance, and thus to provide equal sensitivity to groups across all
redshifts. The linking lengths scale as $D_{\rm L} = D_0 R$ and $V_{\rm L}
= V_0 R$, where $D_0$ and $V_0$ are the linking lengths at some fiducial
recession velocity $V_{\rm f}$, and $R$ is a scaling factor given by:

\begin{equation}
\label{requat}
R = \left[ \frac{\int_{-\infty}^{M_{12}} \phi(M){\rm d}M}
{\int_{-\infty}^{M_{\rm lim}} \phi(M){\rm d}M} \right]^{-1/3},
\end{equation}

\noindent where $M_{\rm lim}$ and $M_{12}$ are the absolute magnitudes of
the faintest galaxy visible at distances of $V_{\rm f}/H_0$ and $V/H_0$
respectively, and $\phi(M)$ is the galaxy luminosity function of the
sample.

In the present analysis, such a scaling factor is in principle unnecessary
because the adoption of a fixed absolute magnitude limit ($M_B < -19$, as
discussed above) removes this distance dependence of the luminosity
function sampling (ie. $M_{\rm lim}$ and $M_{12}$ are both equal to $-19$,
making $R$ unity). However, the 2dFGRS has a slightly varying magnitude
limit, and the spectroscopic completeness varies with position on the
sky. These do require (minor) correction (cf. Merch{\'a}n \& Zandivarez
2002), \nocite{mer02} and can be accounted for by redefining the scaling
factor as

\begin{displaymath}
R = \left[ \frac{C_1 + C_2}{2} \right]^{-1/3},
\end{displaymath}

\noindent where $C_1$ and $C_2$ are the redshift completeness for
magnitudes brighter than the apparent magnitude corresponding to $M_B =
-19$ for the redshift of the galaxies in question, at the location in the
2dFGRS of two galaxies whose linkage is being considered. These redshift
completeness values can be evaluated using the redshift completeness,
limiting magnitude, and $\mu$-masks provided by the 2dFGRS team (see
Colless \etal\ 2001 for details).

The transverse linking length, $D_{\rm L}$, is set by defining groups to be
those regions with a mean galaxy density contrast $\delta \rho / \rho$ in
excess of 80 (cf. Merch{\'a}n \& Zandivarez 2002; Ramella, Pisani \& Geller
1997).  \nocite{mer02,ram97} This relates to $D_0$ through the expression

\begin{displaymath}
\frac{\delta \rho}{\rho} = \frac{3}{4\pi D_0^3}
\left( \int_{-\infty}^{M_{\rm lim}} \phi(M) {\rm d}M \right)^{-1} -1.
\end{displaymath}

\noindent Taking the Schechter fit to the luminosity function of 2dFGRS
galaxies ($\phi^* = 4.6 \times 10^{-3}$Mpc$^{-3}$, $\alpha = -1.21$, $M_B^*
= -20.59$, for $H_0 = 65$\,km\,s$^{-1}$Mpc$^{-1}$; Norberg \etal\
2002)\nocite{nor02}, for an absolute magnitude limit of $M_B = -19$, a
fiducial transverse linking length of $D_0 = 1.25$\,Mpc is derived.

Regarding the velocity linking length, $V_{\rm L}$, this needs to be larger
than the equivalent distance linking length because peculiar velocities
smear galaxies out along the distance axis. As discussed by Nolthenius \&
White \shortcite{nol87}, there is a problem if the standard Huchra \&
Geller approach of scaling the velocity linking length with distance
according to Equation~\ref{requat} is applied to surveys sampling to
moderate redshifts: if a fiducial velocity linking length is chosen which
is large enough to cope with peculiar motions in nearby galaxies, then when
this is scaled with distance it quickly becomes so large that distant
groups are seriously contaminated by foreground and background
galaxies.\footnote{For example, Merch{\'a}n and Zandivarez
\shortcite{mer02} determine a best-fit value of $V_L = 200$\,km\,s$^{-1}$
(at their fiducial recession velocity $V_f = 1000$\,km\,s$^{-1}$) for their
analysis of 2dFGRS groups, by using mock catalogues to investigate the
reliability and completeness of different values. However, using the
scaling of Equation~\ref{requat}, this corresponds to a linking length of
about 900\,km\,s$^{-1}$ for the absolute magnitude limit of $M_B = -19$
used in this paper. This value is clearly too large, and would lead to the
identification of many spurious groups. Indeed, since Merch{\'a}n and
Zandivarez do use this scaling, and study groups out to $z \sim 0.25$,
their linking length is very large at higher redshifts and their group
catalogue is likely to be highly contaminated beyond $z \sim 0.07$.}  The
current analysis avoids this problem, since the adoption of a fixed
absolute magnitude limit essentially removes the requirement of this
scaling. All that is required here is to choose a suitable value of $V_0$.

Nolthenius \& White \shortcite{nol87} find that the balance between
retaining valid group members and minimising contamination from non-members
is optimised for a velocity linking length of between 1.5 and 2 times the
typical velocity dispersion of groups. Merch{\'a}n and Zandivarez
\shortcite{mer02} and Norberg \etal\ \shortcite{nor03} have determined the
average velocity dispersion of galaxy groups in the 2dFGRS to be 250 and
225\,km\,s$^{-1}$, respectively. Therefore, a fiducial velocity linking
length of $V_0 = 450$\,km\,s$^{-1}$ was adopted here.

In this way, the friends--of--friends mechanism was used to link pairs of
galaxies and hence build up a catalogue of groups. 25\% of the $M_B < -19$
galaxies are found to be isolated galaxies (lower than the $\sim 45$\%
fraction that Norberg \etal\ derived for all 2dFGRS galaxies, because more
massive galaxies are preferentially found in groups or clusters), 22\% in
`groups' of 2 or 3 galaxies, a further 28\% in groups of 4-15 galaxies and
25\% in still richer structures. 

A robust estimate of the velocity dispersion of these galaxy groups was
obtained using the `gapper' estimator \cite{wai76} for groups with less
than 15 members and the `biweight' estimator for larger groups (cf. Beers,
Flynn \& Gebhardt 1990; Girardi \etal\ 1993)\nocite{bee90,gir93}. The
virial radius and the mass of the group are then calculated according to
the method of Giraldi \& Giuricin \shortcite{gir00}. The median velocity
dispersion of groups with 4 or more members is 220\,km\,s$^{-1}$, with a
mean radius of 2.3\,Mpc and a mean mass of $5 \times 10^{13}
M_{\odot}$. All of these values are consistent with previous determinations
of the properties of nearby groups.

\section{Results}
\label{results}

\subsection{Galaxy Surface Densities}
\label{surfdens}

Figure~\ref{agnfrac} shows the variation with local projected galaxy
surface density of the fraction of galaxies which host radio sources
associated with AGN activity, and the equivalent fraction which host
star-forming radio sources. The environmental dependence of these two
samples is remarkably different.

The star-forming galaxies show a strong correlation with environment in the
sense that star formation is suppressed in high density environments; this
is the result which is well--known at optical wavelengths, but which is now
independently confirmed using a radio--based study. There is no evidence
for a break in the correlation below $\sim 1$\,Mpc$^{-2}$, as was found by
Lew02; the dependence upon environment continues to the lowest surface
densities.  These results are interpreted in Section~\ref{sfdiscuss}.

The radio--loud AGN fraction shows very little dependence upon environment,
with only the lowest density bin deviating significantly from a flat
distribution.  The surface density bins were defined to contain roughly
equal numbers of AGN, in order to optimize the signal--to--noise; this also
minimises the scatter between the data points a little bit, possibly giving
a overly flat appearance. To test this, a Kolmogorov-Smirnov (KS)
two--population test was applied to the surface density distribution of the
AGN as compared to that of all galaxies. The probability that the AGN are
not simply drawn randomly from all galaxies is 97\% (ie. a $\sim 2.5\sigma$
result). This result is largely driven by the lack of AGN at the lowest
surface densities: the lowest surface density of galaxies around an AGN is
0.08\,Mpc$^{-2}$ whilst over 8\% of the 2dFGRS galaxy population have local
galaxy densities below this value. Considering only galaxies with galaxy
surface densities above 0.1\,Mpc$^{-2}$, the significance of any difference
between the AGN and all galaxies is below $1\sigma$.

\begin{figure}
\centerline{
\psfig{file=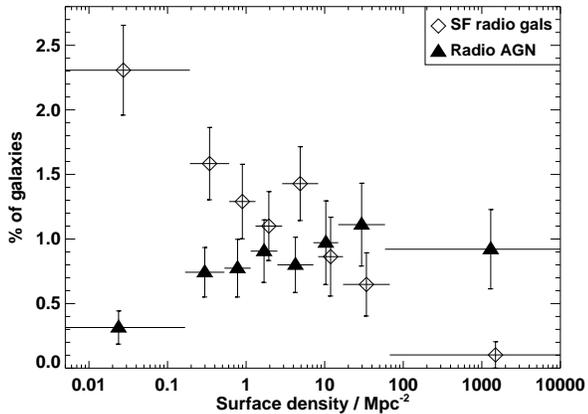,angle=90,width=8.2cm,clip=}
}
\caption{\label{agnfrac} The fraction of galaxies within the 2dFGRS
catalogue that are associated with radio sources securely classified as
either star forming or AGN, as a function of the local projected galaxy
surface density. The error bars plotted (here and in subsequent figures)
correspond to the simple Poissonian uncertainty. The frequency of star
forming radio sources is greatly suppressed in dense environments, whilst
AGN activity is roughly independent of environment, except possibly in the
most rarefied regions. [Note: in this and some subsequent figures, data
points of different samples are offset an equal and opposite small distance
along the x--axis for clarity].}
\end{figure}

\begin{figure}
\centerline{
\psfig{file=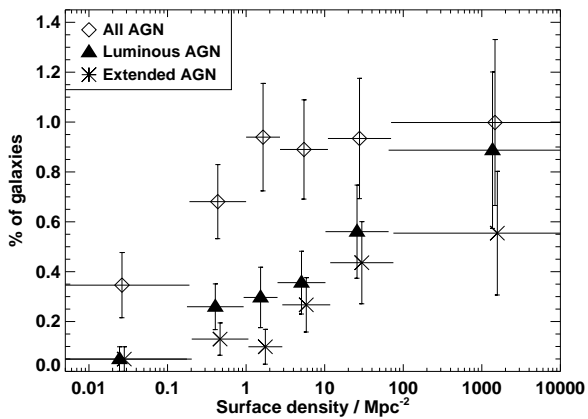,angle=90,width=8.2cm,clip=}
}
\caption{\label{lumagn} The fraction of galaxies within the 2dFGRS
catalogue that are associated with radio-loud AGN, as a function of the
local projected galaxy surface density, compared to the equivalent
fractions considering only luminous AGN (those with 1.4\,GHz radio
luminosities above $10^{23}$W\,Hz$^{-1}$) or only those AGN whose radio
emission is extended in the NVSS data. Both luminous AGN and extended AGN
are significantly more common in richer environments. }
\end{figure}

\begin{figure}
\centerline{
\psfig{file=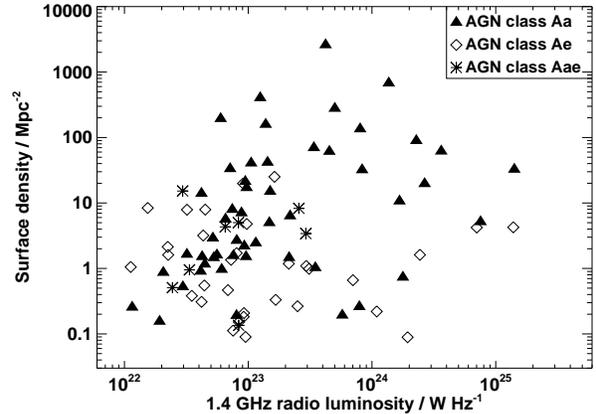,angle=90,width=8.2cm,clip=}
}
\caption{\label{surfrad} The local projected galaxy surface density of AGN
plotted against their 1.4\,GHz radio luminosity, with the three different
symbols showing the three different AGN classes separately. The absorption
line AGN (Aa class) have surface densities strongly correlated ($> 99.9$\%
significance) with radio luminosity, while the two emission line AGN
classes (Ae and Aae) show no such correlation. Indeed, emission line AGN of
all radio luminosities avoid the richest environments. The Aae radio
galaxies all have relatively low radio luminosities.}
\end{figure}

\begin{figure}
\centerline{
\psfig{file=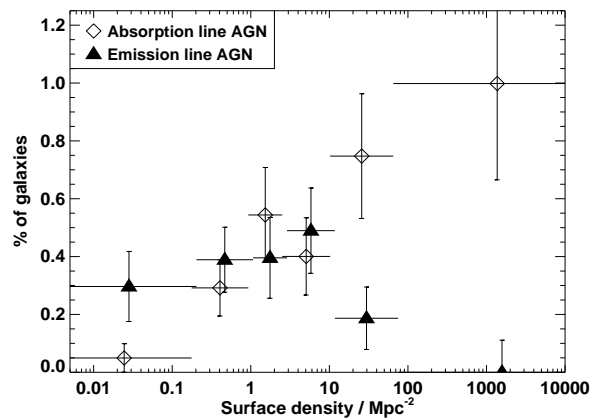,angle=90,width=8.2cm,clip=}
}
\caption{\label{agnclass} The fraction of galaxies within the 2dFGRS
catalogue that are associated with radio-loud AGN as a function of the local
projected galaxy surface density, separated into emission--line AGN (Ae and
Aae classes of Sad02) and absorption line AGN (Aa class). The absorption
line AGN are preferentially located in richer environments, whilst the
emission line AGN avoid dense regions. }
\end{figure}

This apparent lack of any dependence of AGN activity on galaxy surface
density hides a considerable quantity of information on AGN activity that
can be uncovered in more detailed analysis. In Figure~\ref{lumagn} this
dependence of AGN fraction is shown again, and compared to the equivalent
fraction for the two subsets of `Luminous AGN' and `Extended AGN',
described in Section~\ref{radsamp}. The fraction of galaxies hosting AGN
associated with each of these two classes shows a marked increase with
increasing local galaxy density; KS tests show them to be inconsistent
with flat distributions at the 99.5 and 99.7\% significance levels
respectively.

This correlation between AGN luminosity and local environment is perhaps
better illustrated in Figure~\ref{surfrad} which shows the projected
surface density versus the radio luminosity for all of the AGN in the
sample, separated into the three AGN classes defined by Sad02. Overall, a
Spearmann-Rank correlation test indicates that radio luminosity and surface
density are correlated at the 99.6\% significance level. However, the most
striking aspect of this plot is the difference between the properties of
the three classes of AGN. The absorption--line AGN (class Aa) show a strong
correlation between radio luminosity and galaxy surface density ($>99.95$\%
significance, using a Spearmann-Rank correlation test), but the
emission--line AGN (class Ae) have no such correlation: these AGN cover the
full range of radio luminosities, but completely avoid rich environments,
showing the same range of low galaxy surface densities at all radio
luminosities. The Aae class of AGN, which show both absorption features and
weak emission lines, are relatively rare. These occupy the region of the
diagram where the Aa and Ae classes overlap, namely, they are confined to
regions of relatively low radio luminosity and low galaxy density. The
restriction of the Aae class to low radio luminosities may be because at
higher radio luminosities their emission lines would have been stronger
(due to the radio luminosity versus emission line luminosity correlation
for emission line radio sources; cf Rawlings \& Saunders
1991)\nocite{raw91b}, and if these dominated the spectrum the AGN would
have been classified as Ae AGN. Given that the Aae AGN have surface
densities comparable to the Ae class, in the analyses that follow these two
classes are often considered together as emission--line AGN.

The percentage of galaxies in each of the absorption and emission line AGN
classes, as a function of galaxy surface density, is further illustrated in
Figure~\ref{agnclass}: while absorption-line AGN prefer richer
environments, emission line AGN avoid them. The surface density
distributions of the two samples differ at the 99.8\% significance
level. It is tempting to ascribe this difference to differences in the host
galaxies of the AGN: the absorption--line AGN must all be hosted by
elliptical galaxies, and such galaxies are common in cluster environments,
but emission--line AGN could plausibly have a range of host galaxy
types. However, as Figure~\ref{absrad} shows, there are no significant
differences between the emission and absorption--line AGN in terms of their
radio and optical luminosities, suggesting that there are no fundamental
differences in host galaxies. Table~\ref{tabprop} provides mean values for
various properties of these AGN: there is only a 0.24 magnitude ($\lta
2\sigma$) difference in mean host galaxy absolute magnitude between the Aa
and Ae classes, whilst the difference in mean local galaxy surface density
is at the $\gta 5\sigma$ level. Possible explanations for the differences
between the emission and absorption line AGN, and the consequences of these
results, are discussed in Section~\ref{absemis}.

Comparing the properties of the AGN with those of the elliptical galaxy
sample selected from the SDSS data (Table~\ref{tabprop}), the mean local
surface densities are similar. Figure~\ref{agnellip} shows the fraction of
all elliptical galaxies (statistically constructed from the morphological
mix as a function of environment derived from the SDSS subsample) that host
absorption--line AGN as a function of galaxy surface density. Because of
the prevalence of early--type galaxies in clusters this distribution is
flattened from the equivalent fraction in all galaxies, but it is still
evident that absorption--line AGN have a weak preference for richer 
environments, even compared to elliptical galaxies in general.

\begin{figure}
\centerline{
\psfig{file=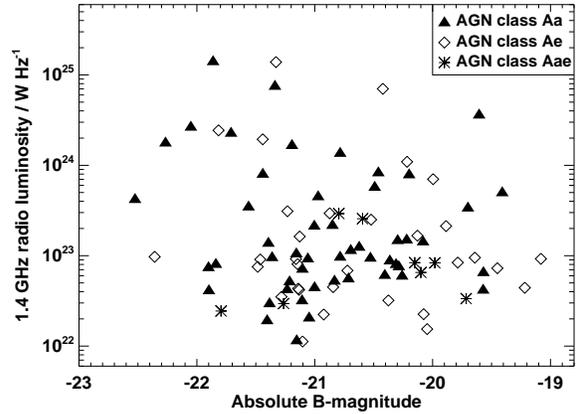,angle=90,width=8.2cm,clip=}
}
\caption{\label{absrad} The B--band absolute magnitude of the AGN plotted
against their 1.4\,GHz radio luminosity, with the three different symbols
showing the three different AGN classes separately. Apart from the Aae
class being confined to low radio luminosities, there are no statistically
significant differences between the host galaxies of the three classes.}
\end{figure}

\begin{table*}
\caption{\label{tabprop} The number of objects in various galaxy and AGN 
samples, together with the mean values of various properties of these
galaxies.}
\begin{tabular}{lcccccc}
\hline
Radio source type & 
N &
z &
log$_{10}L_{\rm 1.4\,GHz}$ &
$M_B$ &
log$_{10}$(Surf Dens) 
\\
&
&
&
W\,Hz$^{-1}$ &
&
Mpc$^{-2}$
\\
\hline
AGN Class Aa   &  51   & 0.073 & $23.25 \pm 0.10$ & $-20.91 \pm 0.10$ & 
$0.87 \pm 0.14$ \\
AGN Class Ae   &  32   & 0.069 & $23.14 \pm 0.13$ & $-20.67 \pm 0.14$ & 
$0.01 \pm 0.12$ \\
AGN Class Aae  &   8   & 0.071 & $22.86 \pm 0.14$ & $-20.55 \pm 0.25$ & 
$0.34 \pm 0.24$ \\   
\hline 
SF galaxies    & 154   & 0.051 & $22.45 \pm 0.03$ & $-20.48 \pm 0.06$ & 
$-0.17\pm 0.07$~~ \\
\hline
All galaxies   & 50684 & 0.076 &     ----         & $-19.99 \pm 0.01$ &  
$0.35 \pm 0.01$ \\
All gals (100k sample) & 21085 & 0.075 &  ----    & $-20.00 \pm 0.01$ &  
$0.26 \pm 0.02$ \\
Early-types    & 3746  & 0.078 &     ----         & $-20.23 \pm 0.01$ & 
$0.82 \pm 0.02$ \\ 
\hline
\end{tabular}
\end{table*}

\begin{figure}
\centerline{
\psfig{file=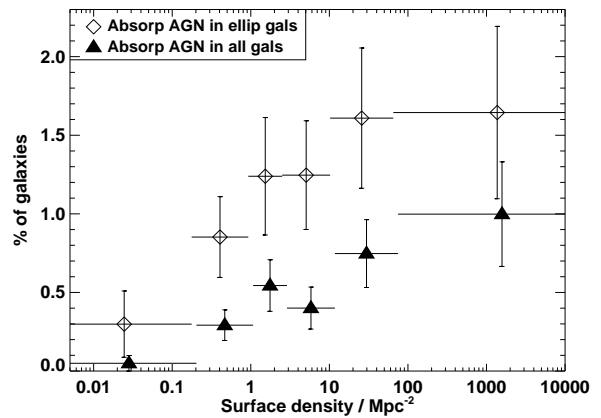,angle=90,width=8.2cm,clip=}
}
\caption{\label{agnellip} The fraction of radio-loud absorption--line AGN
amongst all galaxies within the 2dFGRS catalogue, as a function of the
local projected galaxy surface density, compared against the equivalent
fraction amongst only early--type galaxies (since all the absorption line
AGN will be hosted by early--type galaxies). The prevalence of early-type
galaxies in rich environments means that, although with increasing local
galaxy density an increasing fraction of galaxies host absorption--line
AGN, early--type galaxies host absorption--line AGN with comparable
probabilities across all environments except in the most rarefied
regions. }
\end{figure}

\subsection{Group and Cluster membership}
\label{agnenv}

\begin{figure}
\centerline{
\psfig{file=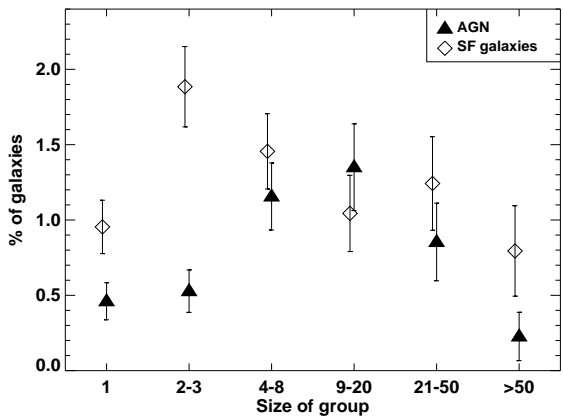,angle=90,width=8.2cm,clip=}
}
\caption{\label{sfgpsize} The fraction of galaxies which host radio--loud
AGN or radio--selected star forming galaxies as a function of the size of
the group (as measured in $M_B < -19$ galaxies) in which the galaxy is
located. Isolated galaxies are rarely luminous star forming galaxies, but
beyond that star forming galaxies show a weak preference to be in the
smallest groups possible. AGN are preferentially found in moderate groups
and poor clusters.}
\end{figure}

\begin{figure}
\centerline{
\psfig{file=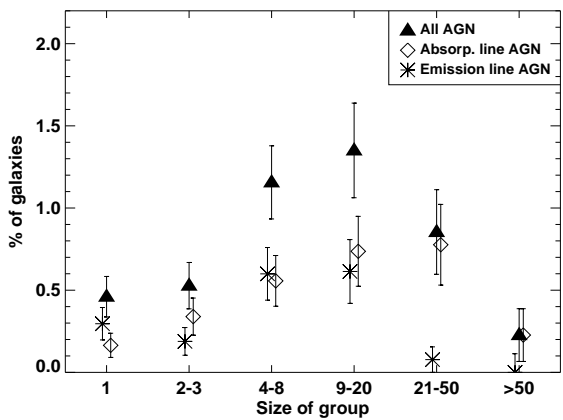,angle=90,width=8.2cm,clip=}
}
\caption{\label{gpsize} The fraction of galaxies which host radio--loud AGN
as a function of the size of the group (as measured in $M_B < -19$
galaxies) in which the galaxy is located, split into subclasses of
absorption-line AGN (Aa class of Sad02) and emission-line AGN (Ae and Aae
classes). The emission--line AGN avoid even moderately rich clusters, while
the absorption--line AGN lie preferentially in poor to moderate clusters,
but are largely absent from rich clusters. }
\end{figure}

\begin{figure}
\centerline{
\psfig{file=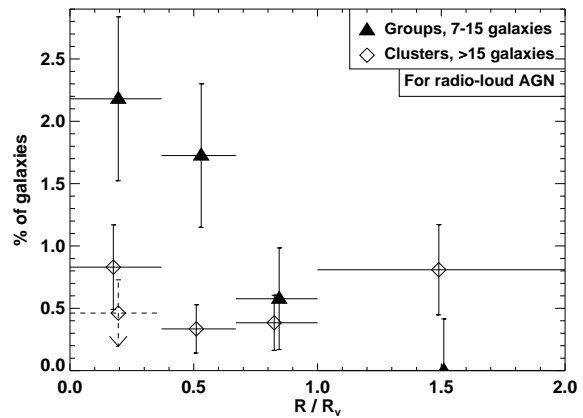,angle=90,width=8.2cm,clip=}
}
\caption{\label{gprad} The radial distribution (in terms of projected
radius of a galaxy from the group\,/\,cluster centre divided by the projected
virial radius of that group\,/\,cluster) of the AGN fraction in both group
(7-15 galaxies) and cluster ($>$15 galaxies) environments. For the inner
radial bin of the cluster AGN fraction, the lower data point represents the
value after correcting for obvious brightest cluster galaxies. AGN are
preferentially found in the inner regions of galaxy groups, but in clusters
they are found at all radii, perhaps even preferentially at larger radii.}
\end{figure}

The fractions of galaxies that host radio--loud AGN activity or
radio--selected star formation activity are shown in Figure~\ref{sfgpsize}
as a function of the size of group or cluster in which that galaxy lies (as
determined by the friends--of--friends analysis; see Section~\ref{gpclus}).
Star forming galaxies prefer to avoid richer environments, although this
dependence is clearly weaker than that on local environment found above.
Star forming galaxies are also relatively rare in isolated galaxies: this
latter result is most likely related to the radio selection, which picks
out only galaxies with relatively high star formation rates ($\gta 5
M_{\odot}$/yr).  AGN are most common in moderate groups and poor
clusters. The fraction of galaxies with AGN activity is lower in relatively
isolated environments (1--3 galaxies) and also in the richest clusters
($>50$ galaxies); a KS test gives a probability of 99.2\% that the AGN are
not simply randomly drawn from all galaxies.

By sub-classifying the AGN into emission--line AGN and absorption--line AGN
(cf. Figure~\ref{gpsize}) it is apparent that the fall-off in the AGN
fraction in rich environments happens earlier for emission--line AGN than
absorption--line AGN. The former are essentially absent from all clusters
of more than 20 galaxies, whilst the latter still strongly populate the
moderate clusters (21-50 galaxies) and are only absent from the very
richest clusters. A KS two--sample test indicates that, overall, the
difference between the group\,/\,cluster environments of the emission and
absorption--line AGN is significant at the 90\% level, but this difference
is concentrated in this single `moderate cluster' environment, in which the
contrast is much more significant (an emission to absorption--line AGN
ratio of 1:10, compared to approximately 1:1 in poorer environments).

The low fraction of galaxies associated with AGN in relatively isolated
environments is in agreement with the results of the local galaxy surface
density analysis, in which a drop in the AGN fraction was also seen at the
lowest surface densities. The decreased fraction of radio--loud AGN in the
richest environments seen in the group membership analysis is, however,
much more pronounced than anything in the galaxy surface densities. This
apparent inconsistency arises because these two analyses sample
environments on very different scales, which are not necessarily highly
correlated: for example, a compact group of 10 galaxies can have a higher
galaxy surface density than even the central regions of a diffuse cluster
of 50 galaxies, while galaxies in a single rich clusters can span a wide
range of local environments from very high galaxy surface densities at the
cluster centre to much lower on the outskirts.

This latter point can be investigated by examining the variation of the AGN
fraction with position in a group or cluster. Figure~\ref{gprad} shows that
in galaxy groups the AGN are predominantly found in the central regions,
whilst for clusters they are found across a wide range of radii. It is
already known that brightest cluster galaxies often host AGN activity, and
after correcting for these AGN (this correction may be even larger than
that indicated on the figure, as only obvious brightest cluster galaxies
were removed) there may even be a weak preference for AGN to lie at large
radii if account is taken that some of the AGN at small projected radii may
be at larger radii in a 3-dimensional analysis.

It is interesting to compare the variations of AGN fraction in
Figures~\ref{agnfrac} and \ref{sfgpsize}: the large--scale environment
(group, cluster, etc) of a galaxy seems to be more important than the
small--scale environment (local galaxy surface density) in determining
whether a galaxy will undergo AGN activity. This result is in contrast to
that of the star formation properties of galaxies, where correlations are
seen with both scales of environment but the stronger correlation is with
the local galaxy density.

\section{Discussion}
\label{discuss}

\subsection{Star forming galaxies}
\label{sfdiscuss}

For radio-selected star forming galaxies a strong correlation is seen
between the fraction of galaxies which are undergoing star formation and
the density of the local environment, in the sense that star formation is
suppressed in higher density environments. This result is equivalent to
that which has been found many times before in optical studies (e.g.
Dressler \etal\ 1985; Balogh \etal\ 1998; Hashimoto \etal\ 1998; Carter
\etal\ 2001; Lew02; Mart{\'\i}nez \etal\ 2002; Gom03; MS03). The fraction
of star forming galaxies is also found to be dependent upon the
large--scale environment (group\,/\,cluster etc), in a similar way.
\nocite{dre85,bal98,has98,car01,Lew02,mar02b,Gom03,mat03}

Lew02 found evidence for a density threshold at a projected galaxy surface
density of $\sim 1$ galaxy per square Mpc, below which star formation is
essentially independent of environment; Gom03 found a similar critical
density threshold, although in this case the break is less clear and some
residual environmental dependence remains at lower galaxy densities. No
such threshold is seen in the current observations. In part this may be
related to the fact that the radio selection only picks out the most
rapidly star forming galaxies: Hashimoto \etal\ \shortcite{has98} suggested
that galaxies with high levels of star formation were very sensitive to
local environment even outside of clusters, whilst those with lower star
formation levels were less sensitive. However, this is unlikely to be the
only effect since MS03 also found that the correlation of star formation
activity with environment held across the full galaxy density range of
their sample (as was also found in previous studies, e.g. Hashimoto \etal\
1998). MS03 proposed three possible reasons to explain the difference
between their results and those of Lew02 and Gom03. Firstly, MS03 restrict
their analysis to field galaxies and so can convert redshifts into
distances and use a volumetric local density estimate, as compared with the
projected surface densities used by Lew02 and Gom03. Secondly, MS03 sample
a magnitude further down the luminosity function than Lew02, and nearly two
magnitudes fainter than Gom03; their sample will therefore include larger
fractions of low luminosity star forming galaxies. Thirdly, the Lew02
results are based upon regions around clusters of galaxies, with the lowest
densities corresponding to the cluster outskirts, and thus may give biased
results with respect to a true field sample.

The results presented here provide some insight as to which of these
factors is truly responsible. The magnitude limit of the current sample is
equivalent to that of Lew02 and, like both Lew02 and Gom03, a projected
surface density is used to estimate the local environment of the galaxies,
yet no evidence for a critical density threshold is found. This suggests
that neither of these two effects is responsible. More likely, the density
threshold in the Lew02 studies arises as a result of restricting the
analysis to only cluster environments. In particular, these authors select
companion galaxies using a cut in velocity of three times the velocity
dispersion of the clusters, as opposed to the fixed velocity width adopted
here. It is not clear whether this technique is really appropriate in the
`field' environments at large cluster radii; the very wide velocity bins
that result may contain large numbers of unrelated galaxies, and therefore
introduce significant scatter into the surface density measurements in
these rarefied environments, washing out any intrinsic correlation.

The results from the radio--selected star--forming galaxy sample support
the idea that local environment plays a very important role in determining
the star formation activity of a galaxy. This effect is still observed at
very low galaxy densities, which suggests that ubiquitous processes such as
gas stripping through tidal interactions and galaxy harassment are more
important than ram pressure and evaporative stripping of gas by a hot
intergalactic medium, which only function in cluster environments.

It is finally worth noting that the population of radio--selected dusty
star-forming galaxies, found in the central regions of clusters by Miller
\& Owen \shortcite{mil02a}, is clearly not a dominant contributant since
the distribution of star forming galaxies selected on their radio emission
(unaffected by dust) is comparable to that of optically--selected samples.

\subsection{AGN Environments}
\label{agndiscuss}

The fraction of galaxies hosting radio--loud AGN is seen to show little
dependence upon the local galaxy surface density except in the very poorest
environments. On the other hand, AGN show a strong preference to be found
in galaxy groups or poor clusters, and tend to avoid both isolated
environments and rich clusters. This latter result spectroscopically
confirms the indications of previous imaging studies of the most powerful
nearby radio sources (e.g. Prestage \& Peacock 1988; Hill \& Lilly
1991)\nocite{pre88,hil91}, and is also in line with the small--scale
spectroscopic survey of Miller \etal\ \shortcite{mil02b}

The AGN fraction appears to be much more strongly dependent upon the
large--scale environment of a galaxy (group\,/\,cluster etc) than upon the
small--scale environment evaluated through the local projected surface
density of galaxies. This result has important consequences for
understanding AGN activity. Radio--loud AGN do not simply trace the
distribution of elliptical galaxies, and therefore the presence of a
supermassive black hole, although essential, is not the principle driver of
AGN activity. Nor is the general availability of cold gas the key factor,
since the dependence of the AGN fraction on environment is radically
different from that of the star forming population. There must be an
alternative mechanism which in some way controls the triggering of the AGN
activity. The environmental dependence of the AGN fraction strongly
supports the argument that galaxy interactions or mergers provide this
mechanism (e.g. Heckman \etal\ 1986)\nocite{hec86}: due to the lower
velocity dispersions of galaxy groups, interactions and especially mergers
are far more common in these environments than in the rich virialised
environments of cluster cores. The observation that the AGN fraction is
highest in the central regions of groups (where the galaxy density is
highest) or the outer regions of clusters (perhaps infalling galaxies, or
galaxy groups, more likely to undergo interactions) provides further
support for this model.

One fundamental issue is whether radio--loud AGN are representative of all
AGN, in which case the results derived in this paper would be applicable to
all AGN samples, or whether the radio selection produces some biased
subsample.  Until it is fully understood what causes an AGN to become
radio--loud, this question cannot be definitively answered. However,
studies of the most powerful radio--loud quasars have found no significant
differences between either the host galaxies or the environments of these
objects and those of radio--quiet quasars selected to have the same optical
luminosities (e.g. McLure \etal\ 1999; Nolan \etal\ 2001; McLure \& Dunlop
2001)\nocite{mcl99,nol01,mcl01}, except that radio--loud quasars are
limited to the absolute upper end of the black hole mass function ($M_{\rm
bh} > 10^9 M_{\odot}$), whilst the radio quiet quasars extend to slightly
lower values ($M_{\rm bh} > 5 \times 10^8 M_{\odot}$; e.g. Dunlop \etal\
2003). This suggests that, at least if analysis is restricted to the most
powerful AGN, the radio--loud AGN are likely to be a reasonably unbiased
subsample. At lower optical and radio luminosities the situation is less
clear.

Given that the AGN fraction seems to depend more strongly on the
large--scale than the local environment of their host galaxy, it will be
interesting to investigate the dependence upon still larger scale. A visual
comparison of the AGN and galaxy group\,/\,cluster distributions suggests
that the AGN trace the large--scale structure of the groups and clusters,
whilst avoiding the centres of the richest clusters; the very patchy
redshift coverage of the 2dFGRS at the time that the Sad02 radio sample was
constructed makes a quantitative analysis infeasible, however.
Qualitatively, it is interesting to note there are already 9 radio--loud
AGN associated with the largest supercluster structure in the 2dFGRS (at RA
13h, Dec -2$^{\circ}$, $z \sim 0.084$, spanning a physical size of about
100\,Mpc east--west), and the very incomplete redshift coverage for this
supercluster when the Sad02 analysis took place suggests that this number
will increase significantly.  This is reminiscent of the recent results of
Brand \etal\ \shortcite{bra03}: these authors used the three--dimensional
distribution of NVSS radio sources, as determined through a spectroscopic
follow--up programme in dedicated sky regions, to discover at least one,
and possibly two, 100\,Mpc--scale superstructures at $z \sim 0.3$ through
radio source overdensities. These results suggest that studies of
overdensities of radio sources in deep surveys may prove to be a powerful
tool for discovering high redshift superclusters. In line with the argument
above that radio--loud AGN are not a highly biased sample of AGN, it is not
only radio sources which could be used for this: optical AGN overdensities
may also trace these large structures (e.g. Williger \etal\ 2002; Haines
\etal\ 2003),\nocite{wil02b,hai03} and especially the peripheral regions of
these (e.g. S\"ochting \etal\ 2002).\nocite{soc02}

\subsection{Absorption versus emission line AGN}
\label{absemis}

One of the most striking results to come out of the 2dFGRS analyses is the
difference between the properties of the emission and absorption line
AGN. Despite having comparable host galaxy properties and covering a
similar range of radio luminosities, the environments of these two AGN
classes differ dramatically. In low density environments there is a roughly
equal split between absorption and emission line AGN, but in the densest
environments essentially none of the AGN show strong line emission. This is
reminiscent of the result of X-ray studies of nearby clusters, which found
that the majority of the cluster X--ray AGN were associated with passive
elliptical galaxies \cite{mar02a}

The difference between emission and absorption--line AGN may be related to
that between `normal, strong emission line' radio galaxies and the class of
`weak emission line' or `low excitation' radio galaxies found within high
radio power samples \cite{hin79,lai94}. In recent years it has been argued
that low-excitation radio galaxies do not partake in the orientation--based
unified schemes for radio sources (e.g. Barthel 1989)\nocite{bar89}, in
which their stronger emission line counterparts are believed to be drawn
from the same parent population as radio--loud quasars, but oriented such
that direct quasar light is obscured by a torus of dusty material partially
surrounding the nucleus. Harvanek \etal\ \shortcite{har01} recently used
galaxy number count analyses to show that radio galaxies at $z \sim 0.3$
typically live in richer environments than radio--loud quasars, and
suggested that this might pose a problem for unified schemes. Hardcastle
\shortcite{har03} subsequently showed that this was not the case: if the
low excitation radio galaxies are removed from the analysis, then there is
no significant difference between the environments of strong emission line
radio galaxies and those of the radio loud quasars. The entire difference
in the Harvanek \etal\ \shortcite{har01} study was driven by the low
excitation radio galaxies, essentially all of which are found to reside in
much richer environments than those of the strong emission line radio
galaxies and quasars. This result exactly mirrors that derived in this
paper, and suggests that these low excitation radio sources are simply
higher radio power examples of the radio sources classified here as
absorption--line AGN; the correlation between radio power and environmental
richness found in this paper naturally explains why these high radio power
low excitation radio sources typically live in very rich environments.

A key question is what drives this difference between the emission and
absorption line AGN?  Are these fundamentally different types of AGN, or
does the surrounding environment influence the host galaxy to such an
extent that different AGN properties are seen? Although fundamental
differences in the AGN properties cannot be ruled out, the similarity of
the host galaxies of the two different classes makes this hypothesis
difficult to explain, whilst the alternative might be plausible through one
or both of two different mechanisms.

Rawlings \& Saunders \shortcite{raw91b} showed that for the strong emission
line radio sources, there is a tight correlation between the jet power of
the radio source and the emission line luminosity. Barthel \& Arnaud
\shortcite{bar96a} showed that radio sources in clusters have significantly
more luminous radio emission than would be expected from their
far--infrared luminosities, and argued that this was because the confining
effect of the dense intracluster medium reduces adiabatic expansion losses
in the radio lobes and therefore boosts the radio synchrotron emission (see
Barthel \& Arnaud 1996, and references therein). The combination of these
two effects suggests that some of the high radio power absorption line
radio sources, in the denser environments, may be intrinsically lower
jet--power~AGN (with consequently lower expected emission line
luminosities) whose location in a rich cluster has led to boosting of their
radio emission giving rise to their comparatively high radio luminosities.

At first sight, this idea of boosted radio luminosities in cluster
environments appears to be in contrast to indications that there is little
difference between the radio luminosity functions inside and outside of
clusters (e.g. Ledlow \& Owen 1996, and references therein).\nocite{led96}
However, it is important to realise that the radio luminosity function
arises through a complicated combination of the distribution of galaxy
masses (since more luminous galaxies are more likely to host radio sources;
e.g. Sadler \etal\ 1989), the triggering rate of radio sources, and the
ultimate luminosities of those radio sources. A comparable radio luminosity
function in clusters to that in the field may be obtained if fewer radio
sources of a given intrinsic power are triggered in clusters (as suggested
by the results of this paper), but each such radio source has higher radio
luminosity.  Clearly this issue of radio luminosity boosting is one which
requires more detailed study.

An alternative explanation is that the absorption--line radio galaxies show
little line emission, not because their AGN power is low, but because there
is little gas in their surroundings capable of producing line emission. As
has already been discussed for the analysis of star formation activity, a
number of different physical mechanisms act to remove cool gas from
galaxies in the vicinity of clusters, including tidal interactions, galaxy
harassment, ram pressure stripping and evaporative stripping of gas by the
hot intergalactic medium. Radio galaxies in cluster environments are
therefore likely to be gas poor; there must be some cool gas supply
available in order to fuel the central AGN, but this does not need to
extend over the entire galaxy. If there is little cool gas in the galaxy,
it follows directly that the covering fraction of these gas clouds will be
low, and only a tiny proportion of the ionising photons emitted by the AGN
will be intercepted, leading to very low emission line luminosities even
for a relatively powerful AGN.

It seems probable that both of these mechanisms are important at some
level. The first is able to explain the correlation between radio
power and environment, whilst the second can explain why essentially all
AGN in clusters show little or no line emission: boosting of the radio
power alone cannot explain why X-ray sources in clusters also show a lack
of line emission (cf. Martini \etal\ 2002)\nocite{mar02a}. As an aside, it
is worth noting that at high redshifts, powerful radio sources (with strong
emission lines) are often found to lie in cluster environments, indicating
that the separation between absorption and emission line AGN in rich
environments breaks down at earlier cosmic epochs. This is likely to be
because of the much greater availability of gas at these earlier epochs,
even in young cluster environments.

Finally, this striking change in the ratio of absorption to emission line
AGN in rich environments has important consequences for optically--based
studies of AGN environments. In these studies AGN are selected on their
emission--line properties, and so if this result for radio--loud AGN also
holds for radio--quiet AGN, simple examination of the optical spectra will
miss essentially all of the AGN in rich environments. Detailed modelling of
the stellar populations of the galaxies, in order to recover weak emission
lines within deep stellar absorption features, can help significantly here
(Kauffmann et~al, private communication), recovering weak emission line
AGN.  However, if the emission lines are significantly weaker or absent due
to the lack of gas in cluster galaxies, then optical selection will miss
some or many cluster AGN. This would distort any environmental analyses of
optically--selected AGN, and may also introduce significant biases into
other studies of AGN selected by these means.

\section{Conclusions}
\label{concs}

The results of this paper can be summarised as follows:

\begin{itemize}
\item The proportion of radio--selected star forming galaxies decreases
strongly with increasing local galaxy surface density, in the same manner
as found in optical studies of star forming galaxies. This correlation
extends over the full range of galaxy surface densities, with no evidence
for a lower density threshold.

\item Radio--loud AGN activity shows little dependence on local galaxy
surface density, except at the very lowest surface densities where little
AGN activity is found. The larger scale environment is more important in
determining AGN activity: AGN are preferentially found in moderate groups
and poor clusters.

\item The AGN activity traces neither the distribution of galaxy bulges nor
the availability of cold gas in galaxies, meaning that an external
influence is required to trigger the activity. The higher AGN fraction in
environments where conditions are optimised for galaxy interactions and
mergers indicates that these are likely to be an important mechanism.

\item Where AGN are found in poor or moderate richness clusters they are
almost invariably absorption--line AGN, and have relatively high radio
luminosities. This likely reflects the lack of cool gas typically available
for ionisation in cluster environments, and suggests that the radio
luminosity of these sources may have been boosted by their dense
surrounding environment.

\item The substantial drop in the ratio of emission--line to
absorption--line AGN in dense environments implies that, at very least,
considerable care must be taken in selecting samples of AGN from their
optical emission--line properties. Potentially these samples could miss a
large fraction of cluster AGN, in which case results from AGN environmental
studies based upon optically--selected AGN samples would be unreliable.
\end{itemize}

An investigation of the environments of radio--selected AGN over a much
larger area, such as that which will ultimately be possible using the SDSS,
will permit these environmental variations to be studied to a much greater
degree, using a radio sample of sufficient size for more detailed
statistical investigation of AGN subsamples. In addition, such studies
would enable a detailed comparison between optically and radio selected AGN
samples; this is the key to understanding the potential biases of each
method.

It is also important to investigate more powerful AGN than those studied
here: the most powerful radio source in the current study has a 1.4\,GHz
radio luminosity of $1.4 \times 10^{25}$W\,Hz$^{-1}$, while the most
powerful nearby sources have radio luminosities of
$10^{26-27}$W\,Hz$^{-1}$, much more comparable to those found at higher
redshifts. Studies of these sources would permit an investigation of the
cosmic evolution of AGN environments, but such powerful AGN are very rare
in the nearby Universe, and a dedicated redshift survey of their
environments will be required to achieve this.

\section*{Acknowledgements} 

The author would like to thank the Royal Society for financial support
through its University Research Fellowship scheme, Guinevere Kauffmann for
detailed discussions of optical versus radio selection of AGN, and Ignas
Snellen for a number of useful discussions. The author is also grateful to
the referee for prompt and helpful comments on the original version of the
manuscript.  This research has made use of the 2dFGRS, the data for which
has been reduced and released by the 2dFGRS team, to whom the author is
grateful. A modified version of the 2dFGRS mask software of Peder Norberg
and Shaun Cole was also used. The research makes use of the SDSS Archive,
funding for the creation and distribution of which was provided by the
Alfred P. Sloan Foundation, the Participating Institutions, the National
Aeronautics and Space Administration, the National Science Foundation, the
U.S. Department of Energy, the Japanese Monbukagakusho, and the Max Planck
Society.

\bibliography{pnb} 
\bibliographystyle{mn} 

\label{lastpage}
\end{document}